\documentclass[reprint,nofootinbib,showpacs,tightenlines]{revtex4}
\usepackage{graphicx}  
\usepackage{dcolumn}   
\usepackage{bm}        
\usepackage{amssymb}

\begin{document}

\title{Exploring $R_D$, $R_{D^{\ast}}$ and $R_{J/\Psi}$ anomalies}
\author{Rupak~Dutta${}$}
\email{rupak@phy.nits.ac.in}
\affiliation{
${}$National Institute of Technology Silchar, Silchar 788010, India\\
}

\begin{abstract}
Deviations from the standard model predictions have been reported in various observables concerned with the lepton flavor universality. 
At present, the deviation of the measured values of $R_D$ and $R_{D^{\ast}}$ from the standard model expectation is exceeded by 
$2.3\sigma$ and $3.4\sigma$, respectively. 
Very recently LHCb has measured the ratio of branching ratio $R_{J/\Psi} = 
\mathcal B(B_c \to J/\Psi\tau\nu)/\mathcal B(B_c \to J/\Psi\,l\,\nu)$, where $l \in (e,\,\mu)$, to be $0.71\pm0.17\pm 0.18$ which is 
at more than $2\sigma$ away from the standard model prediction. We investigate the anomalies in $R_D$, $R_{D^{\ast}}$, 
and $R_{J/\Psi}$ using a model independent framework with minimal number of new physics couplings. We find various new physics models 
that can explain these anomalies within $1\sigma$. 
\end{abstract}

\maketitle

\section{Introduction}
\label{int}
Lepton flavor universality violation has been the center of attention due to the long standing anomalies that persisted in the ratio
of branching ratios $R_D$ and $R_{D^{\ast}}$, where
\begin{eqnarray}
&&R_{D^{(\ast)}}=\frac{\mathcal B(B \to D^{(\ast)}\tau\nu)}{\mathcal B(B \to D^{(\ast)}l\nu)}\,, \qquad
l \in (e,\,\mu)\,.
\end{eqnarray}
Unlike the individual branching ratio of these decay modes, $R_D$ and $R_{D^{\ast}}$ 
do not suffer from the uncertainties coming from the Cabbibo-Kobayashi-Mashakawa~(CKM) matrix elements and the meson to meson form factors. 
The dependency on the CKM matrix elements exactly cancels in these ratios. Similarly, the uncertainties due to the form factors also 
largely cancel in these ratios and a clean prediction of $R_D$ and $R_{D^{\ast}}$ can be made within the standard model~(SM). Hence, 
any deviation from the SM prediction would clearly indicate the presence
of new physics~(NP). At present, combining the results of $R_D$ and $R_{D^{\ast}}$ measured by 
various experiments such as
BABAR~\cite{Lees:2012xj,Lees:2013uzd}, BELLE~\cite{Huschle:2015rga,Sato:2016svk,Hirose:2016wfn}, and 
LHCb~\cite{Aaij:2015yra,Aaij:2017uff}, i.e, $R_D = 0.407 \pm 0.039 \pm 0.024$ and $R_{D^{\ast}} = 0.304 \pm 0.013 \pm 0.007$ exceed the 
SM predictions by $2.3\sigma$ and $3.4\sigma$, respectively. Again, including the $R_D-R_{D^{\ast}}$ correlation, 
the discrepancy with SM prediction~\cite{Lattice:2015rga,Na:2015kha,Aoki:2016frl,Bigi:2016mdz,Fajfer:2012vx} currently stands at 
about $4.1\sigma$~\cite{hflag}.  
Recently, LHCb~\cite{LHCb,LHCb2} has measured the value of the ratio of branching ratio 
\begin{eqnarray}
R_{J/\Psi}=\frac{\mathcal B(B_c \to J/\Psi\,\tau\nu)}{\mathcal B(B_c \to J/\Psi\,l\nu)} = 0.71\pm 0.17\pm 0.18\,,
\end{eqnarray}
where $l$ is either an electron or muon,  which is at more than 
$2\sigma$ away from the standard model~(SM) prediction~\cite{Wen-Fei:2013uea,Dutta:2017xmj}. In Ref.~\cite{Hirose:2016wfn}, 
the $\tau$ polarization fraction in $B \to D^{\ast}\tau\nu$ decays has also been measured by BELLE Collaboration and it is reported to be 
$P_{\tau}^{D^{\ast}} = -0.38\pm 0.51^{+0.21}_{-0.16}$. All these measurements indicate an upward deviation from 
the SM expectation.

Various model-independent and model-dependent~\cite{Fajfer1,Tanaka:1994ay,
Nierste:2008qe,miki,Crivellin,datta,datta1,datta2,fazio,Celis,He,dutta,Deshpand:2016cpw,Li:2016vvp,Bardhan:2016uhr,Alok:2016qyh,
Ivanov:2015tru,Ivanov:2016qtw,Nandi:2016wlp,Dutta:2016eml,Alonso:2016oyd,Celis:2016azn,Altmannshofer:2017poe,Iguro:2017ysu} 
approaches have been 
carried out to explore NP effects in $R_D$ and $R_{D^{\ast}}$.
Very recently, NP effects on $B_c \to J/\Psi\,\tau\nu$ decays has also been studied by various 
authors~\cite{Dutta:2017xmj,Watanabe:2017mip,Chauhan:2017uil}. 
In view of the new measurement of $R_{J/\Psi}$ made by LHCb, we investigate $R_D$, $R_{D^{\ast}}$, and $R_{J/\Psi}$
anomalies within a model independent framework with minimal number of NP couplings. We use the most general effective Lagrangian in the 
presence of NP which is valid at renormalization scale $\mu = m_b$ and seek to find the minimal number of NP couplings that best fit the 
data. We have considered total $55$ NP 
scenarios based on NP contributions from single operators as well as from two different operators and try to find the scenario that best
explains $R_D$, $R_{D^{\ast}}$, and $R_{J/\Psi}$ anomalies. Although this is not entirely a new idea, most efforts at
explaining these anomalies consider only a subset of these $55$ NP structures.
We do not propose any NP model that can generate such NP structures in particular.
Rather we study it from a purely phenomenological point of view. 
It was shown in Ref.~\cite{Alonso:2016oyd} that lifetime of $B_c$ meson put a severe constraint on scalar type NP interactions. 
Based on various SM 
calculations~\cite{Bigi:1995fs,Beneke:1996xe,Chang:2000ac}, it was inferred that $\mathcal B(B_c \to \tau\nu) \le 5\%$ is necessary 
to comply with the current world average of the $B_c$ lifetime.  
However, the constraint can be relaxed up 
to $30\%$ depending on the value of the total decay width of $B_c$ meson that is used as input for the
SM calculation of various partonic transitions. Very recently, in Ref.~\cite{Akeroyd:2017mhr}, a more significant bound of 
$\mathcal B(B_c \to \tau\nu) \le 10\%$ was obtained by taking the LEP data at the $Z$ peak. Our analysis also takes into account the
indirect constraint coming from $\mathcal B(B_c \to \tau\nu)$ to rule out various NP scenarios. We, however, have not included the 
measured value of $P_{\tau}^{D^{\ast}}$ reported by Belle in our fitting method since the uncertainties associated with it is rather large.

The paper is organized as follows. In section.~\ref{ehha}, we present the most general effective weak Lagrangian for the $b \to c\tau\nu$
quark level transitions in the presence of NP valid at renormalization scale $\mu = m_b$. We also present all the relevant formulas
pertinent for our analysis. The results of our analysis are reported in section.~\ref{rd}. We conclude with a brief summary of our results
in section.~\ref{con}.

\section{Theory framework}
\label{ehha}
We begin with the most general effective Lagrangian for the $b \to c\,l\nu$ quark level transitions in the presence of NP. 
That is~\cite{Bhattacharya, Cirigliano}
\begin{eqnarray}
\mathcal L_{\rm eff} &=&
-\frac{4\,G_F}{\sqrt{2}}\,V_{c b}\,\Bigg\{(1 + V_L)\,\bar{l}_L\,\gamma_{\mu}\,\nu_L\,\bar{c}_L\,\gamma^{\mu}\,b_L + 
V_R\,\bar{l}_L\,\gamma_{\mu}\,\nu_L\,\bar{c}_R\,\gamma^{\mu}\,b_R 
+
\widetilde{V}_L\,\bar{l}_R\,\gamma_{\mu}\,\nu_R\,\bar{c}_L\,\gamma^{\mu}\,b_L 
+
\widetilde{V}_R\,\bar{l}_R\,\gamma_{\mu}\,\nu_R\,\bar{c}_R\,\gamma^{\mu}\,b_R + \nonumber \\
&&S_L\,\bar{l}_R\,\nu_L\,\bar{c}_R\,b_L + 
S_R\,\bar{l}_R\,\nu_L\,\bar{c}_L\,b_R +
\widetilde{S}_L\,\bar{l}_L\,\nu_R\,\bar{c}_R\,b_L +
\widetilde{S}_R\,\bar{l}_L\,\nu_R\,\bar{c}_L\,b_R 
+ 
T_L\,\bar{l}_R\,\sigma_{\mu\nu}\,\nu_L\,\bar{q^\prime}_R\,\sigma^{\mu\nu}\,b_L + \nonumber \\
&&\widetilde{T}_L\,\bar{l}_L\,\sigma_{\mu\nu}\,\nu_R\,\bar{q^\prime}_L\,\sigma^{\mu\nu}\,b_R\Bigg\}
 + {\rm H.c.}\,,
\end{eqnarray}
where $(V_L,\,V_R,\,S_L,\,S_R,\,T_L)$ represents NP couplings that involve left handed neutrino interactions and 
$(\widetilde{V}_L,\,\widetilde{V}_R,\,\widetilde{S}_L,\,\widetilde{S}_R,\,\widetilde{T}_L)$ represents NP couplings that involve right handed 
neutrino interactions. We consider all the NP couplings to be real.
In the presence of such NP, the three body differential branching ratios for $B_q \to (P,\,V)\,l\nu$ decays, where $P(V)$ represents
pseudoscalar~(vector) meson, can be written as~\cite{dutta,Bardhan:2016uhr,Watanabe:2017mip}
\begin{small}
\begin{eqnarray}
&&\frac{d\Gamma^P}{dq^2}(+) =
\frac{8\,N\,|\overrightarrow{p}_P|\,}{3}\Bigg\{\frac{m_l^2}{q^2}\Big[\widetilde{H}_{T0}^2 + \frac{1}{2}\,H_{0T}^2 + \frac{3}{2}\,
H_{tS}^2\Big]\Bigg\}\,,\qquad
\frac{d\Gamma^P}{dq^2}(-) =
\frac{8\,N\,|\overrightarrow{p}_P|\,}{3}\Bigg\{\frac{m_l^2}{q^2}\Big[H_{T0}^2 + \frac{1}{2}\,\widetilde{H}_{0T}^2 + \frac{3}{2}\,
\widetilde{H}_{tS}^2\Big]\Bigg\}\,,
\end{eqnarray}
\end{small}
where
\begin{small}
\begin{eqnarray}
&&H_0 = \frac{2\,m_{B_q}\,|\overrightarrow{p}_P|}{\sqrt{q^2}}\,F_{+}(q^2)\,,\qquad
H_t = \frac{m_{B_q}^2 - m_P^2}{\sqrt{q^2}}\,F_0(q^2)\,,\qquad
H_S=\frac{m_{B_q}^2 - m_P^2}{m_b(\mu) - m_{c}(\mu)}\,F_0(q^2)\,,\qquad
H_T=\frac{8\,m_{B_q}|\overrightarrow{p}_P|}{m_{B_q} + m_P}\,F_T\,, \nonumber \\
&&H_{T0}=H_T\,T_L + \frac{\sqrt{q^2}}{m_l}\,H_0\,G_V\,, \qquad
\widetilde{H}_{T0}=H_T\,\widetilde{T}_L + \frac{\sqrt{q^2}}{m_l}\,H_0\,\widetilde{G}_V\,,\qquad
H_{0T}=H_0\,G_V + \frac{\sqrt{q^2}}{m_l}\,H_T\,T_L\,, \nonumber \\
&&\widetilde{H}_{0T}=H_0\,\widetilde{G}_V + \frac{\sqrt{q^2}}{m_l}\,H_T\,\widetilde{T}_L\,,\qquad
H_{tS} = H_t\,G_V + \frac{\sqrt{q^2}}{m_l}\,H_S\,G_S\,, \qquad
\widetilde{H}_{tS} = H_t\,\widetilde{G}_V + \frac{\sqrt{q^2}}{m_l}\,H_S\,\widetilde{G}_S\,.
\end{eqnarray}
\end{small}
Similarly
\begin{small}
\begin{eqnarray}
&&\frac{d\Gamma^V}{dq^2}(+) =
\frac{8\,N\,|\overrightarrow{p}_P|\,}{3}\Bigg\{\frac{m_l^2}{q^2}\Big[\widetilde{\mathcal A}_{TAV}^2 + \frac{1}{2}\,\mathcal A^2_{AVT} +
\frac{3}{2}\,\mathcal A_{tP}^2\Big]\Bigg\}\,,\qquad
\frac{d\Gamma^V}{dq^2}(-) =
\frac{8\,N\,|\overrightarrow{p}_P|\,}{3}\Bigg\{\frac{m_l^2}{q^2}\Big[\mathcal A_{TAV}^2 + \frac{1}{2}\,\widetilde{\mathcal A}^2_{AVT} +
\frac{3}{2}\,\widetilde{\mathcal A}_{tP}^2\Big]\Bigg\}\,,
\end{eqnarray}
\end{small}
where
\begin{small}
\begin{eqnarray}
&&\mathcal A^2_{TAV} = \Big(\mathcal A_{T_0}\,T_L + \frac{\sqrt{q^2}}{m_l}\,\mathcal A_0\,G_A\Big)^2 + 
\Big(\mathcal A_{T_2}\,T_L + \frac{\sqrt{q^2}}{m_l}\,\mathcal A_{||}\,G_A\Big)^2 +
\Big(\mathcal A_{T_1}\,T_L + \frac{\sqrt{q^2}}{m_l}\,\mathcal A_{\perp}\,G_V\Big)^2 \,, \nonumber \\
&&\widetilde{\mathcal A}^2_{TAV} = \Big(\mathcal A_{T_0}\,\widetilde{T}_L + \frac{\sqrt{q^2}}{m_l}\,\mathcal A_0\,\widetilde{G}_A\Big)^2 + 
\Big(\mathcal A_{T_2}\,\widetilde{T}_L + \frac{\sqrt{q^2}}{m_l}\,\mathcal A_{||}\,\widetilde{G}_A\Big)^2 +
\Big(\mathcal A_{T_1}\,\widetilde{T}_L + \frac{\sqrt{q^2}}{m_l}\,\mathcal A_{\perp}\,\widetilde{G}_V\Big)^2 \,, \nonumber \\
&&\mathcal A^2_{AVT} = \Big(\mathcal A_0\,G_A + \frac{\sqrt{q^2}}{m_l}\,\mathcal A_{T_0}\,T_L\Big)^2 +
\Big(\mathcal A_{||}\,G_A + \frac{\sqrt{q^2}}{m_l}\,\mathcal A_{T_2}\,T_L\Big)^2 +
\Big(\mathcal A_{\perp}\,G_V + \frac{\sqrt{q^2}}{m_l}\,\mathcal A_{T_1}\,T_L\Big)^2\,, \nonumber \\
&&\widetilde{\mathcal A}^2_{AVT} = \Big(\mathcal A_0\,\widetilde{G}_A + \frac{\sqrt{q^2}}{m_l}\,\mathcal A_{T_0}\,\widetilde{T}_L\Big)^2 +
\Big(\mathcal A_{||}\,\widetilde{G}_A + \frac{\sqrt{q^2}}{m_l}\,\mathcal A_{T_2}\,\widetilde{T}_L\Big)^2 +
\Big(\mathcal A_{\perp}\,\widetilde{G}_V + \frac{\sqrt{q^2}}{m_l}\,\mathcal A_{T_1}\,\widetilde{T}_L\Big)^2\,,\nonumber \\
&&\mathcal{A}_{tP}=\mathcal{A}_t\,G_A + \frac{\sqrt{q^2}}{m_l}\,\mathcal{A}_P\,G_P \,,\qquad\qquad
\mathcal{\widetilde{A}}_{tP}=\mathcal{A}_t\,\widetilde{G}_A + \frac{\sqrt{q^2}}{m_l}\,\mathcal{A}_P\,\widetilde{G}_P
\end{eqnarray}
\end{small}
and
\begin{small}
\begin{eqnarray}
&&\mathcal{A}_\parallel=\frac{2(m_{B_q}+m_V)A_1(q^2)}{\sqrt 2}\,,\qquad
\mathcal{A}_\perp=-\frac{4m_{B_q}V(q^2)|\vec p_V|}{\sqrt{2}(m_{B_q}+m_V)}\,,\qquad
\mathcal A_{T_1}=-\frac{8\sqrt{2}\,m_{B_q}\,|\overrightarrow{p}_V|}{\sqrt{q^2}}\,T_1\,, \nonumber \\
&&\mathcal{A}_t=\frac{2m_{B_q}|\vec p_V|A_0(q^2)}{\sqrt {q^2}}\,,\qquad
\mathcal{A}_P=-\frac{2m_{B_q}|\vec p_V|A_0(q^2)}{(m_b(\mu)+m_c(\mu))}\,, \qquad
\mathcal A_{T_2}=-\frac{4\sqrt{2}(m_{B_q}^2 - m_{V}^2)}{\sqrt{q^2}}\,T_2\,, \nonumber \\
&&\mathcal{A}_0=\frac{1}{2\,m_V\,\sqrt{q^2}}\Big[\Big(\,m_{B_q}^2-m_V^2-q^2\Big)(m_{B_q}+m_V)A_1(q^2)\,-\,\frac{4M_B^2|\vec p_V|^2}{m_{B_q}+m_V}A_2(q^2)
\Big]\,, \nonumber\\
&&\mathcal A_{T_0} = \frac{2}{m_V}\,\Big[-\Big(m_{B_q}^2 + 3\,m_V^2-q^2\Big)\,T_2(q^2) + \frac{2\,m_{B_q}\,|\overrightarrow{p}_V|}
{m_{B_q}^2 - m_V^2}\,T_3(q^2)\Big]\,.
\end{eqnarray}
\end{small}
Here we denote $G_V = 1+V_L+V_R$, $G_A =1+ V_L - V_R$, $G_S = S_L+S_R$, $G_P=S_L-S_R$, $\widetilde{G}_V=\widetilde{V}_L + \widetilde{V}_R$,
$\widetilde{G}_A=\widetilde{V}_L - \widetilde{V}_R$, $\widetilde{G}_S=\widetilde{S}_L + \widetilde{S}_R$, and 
$\widetilde{G}_P=\widetilde{S}_L - \widetilde{S}_R$. Again, $|\overrightarrow{p}_{P(V)}| = 
\sqrt{\lambda(m_{B_q}^2,\,m_{P(V)}^2,\,q^2)}/2\,m_{B_q}$ denotes the three momentum vector of the 
outgoing meson. The ratio of branching ratios and $\tau$ polarization
fractions for these decay modes are
\begin{small}
\begin{eqnarray}
&&R_{M} = \frac{\mathcal B(B_q \to M\tau\nu)}{\mathcal B(B_q \to M\,l\nu)}\,,\qquad
P_{\tau}^{M} = \frac{\Gamma^{M}(+) - \Gamma^{M}(-)}{\Gamma^{M}(+) + \Gamma^{M}(-)}\,, 
\end{eqnarray}
\end{small}
where, $l$ is either an electron or a muon and $B_q$ is either a $B$ meson or a $B_c$ meson. Similarly, $M$ refers to the outgoing 
pseudoscalar or vector meson. Again, 
$\Gamma(+)$ and $\Gamma(-)$ denote the decay widths of 
positive and negative helicity $\tau$ lepton, respectively.
\section{Numerical analyses}
\label{rd}
For the numerical estimates of all the observables we first report all the input parameters.  
For the quark, lepton, and meson masses, we use $m_b(m_b)=4.18\,{\rm GeV}$, $m_c(m_b)=0.91\,{\rm GeV}$, $m_e= 0.510998928\times 10^{-3}
\,{\rm GeV}$, $m_{\mu}= 0.10565837151\,{\rm GeV}$, $m_{\tau}= 1.77682\,{\rm GeV}$, $m_{J/\Psi}= 3.0969{\rm GeV}$, 
$m_{B^-}= 5.27931\,{\rm GeV}$, $m_{B_c}= 6.2751\,{\rm GeV}$,
$m_{D^0}= 1.86483\,{\rm GeV}$, and $m_{{D^{\ast}}^0}= 2.00685\,{\rm GeV}$~\cite{pdg}.
Similarly, for the mean lifetime of $B^-$ and $B_c$ meson, we use $\tau_{B^-} = 1.638 \times 10^{-12}\,{\rm s}$ and 
$\tau_{B_c}= 0.507 \times 10^{-12}\,{\rm s}$~\cite{pdg}. We use $f_{B_c}= 0.434(0.015)\,{\rm GeV}$ from Ref.~\cite{Colquhoun:2015oha}.
The value of CKM matrix element is taken to be $V_{cb}= 0.0409(0.0011)$~\cite{pdg}.   
The uncertainty associated with $f_{B_c}$ and $V_{cb}$ are indicated by the number in parentheses. 

In order to compute the branching fractions and other observables, we need information on various hadronic form factors that parametrizes
the hadronic matrix elements of vector, axial vector, scalar, pseudoscalar, and tensor currents between two mesons.
For the $B_c \to J/\Psi$ hadronic form factors, we follow ref.~\cite{Wen-Fei:2013uea}.
The relevant formula for $V(q^2)$, $A_0(q^2)$, $A_1(q^2)$, and $A_2(q^2)$ pertinent for our
discussion, taken from ref.~\cite{Wen-Fei:2013uea} is
\begin{eqnarray}
F(q^2) = F(0)\,\exp\Big[a\,q^2 + b\,(q^2)^2\Big]\,,
\end{eqnarray}
where $F$ stands for the form factors $V$, $A_0$, $A_1$, and $A_2$ and $a$, $b$ are                              
the fitted parameters.
The numerical values of $B_c \to J/\Psi$ form factors at $q^2 = 0$ and their fitted parameters $a$ and $b$, 
calculated in perturbative QCD~(PQCD) approach, are collected from ref.~\cite{Wen-Fei:2013uea}. Similarly for the $B_c \to J/\Psi$
tensor form factors, we follow Ref.~\cite{Watanabe:2017mip}. The relevant formulas pertinent for our numerical computation are
\begin{eqnarray}
&&T_1(q^2) = \frac{m_b + m_c}{m_{B_c} + m_{J/\Psi}}\,V(q^2)\,,\qquad
T_2(q^2) = \frac{m_b - m_c}{m_{B_c} - m_{J/\Psi}}\,A_1(q^2)\,,\nonumber \\
&&T_3(q^2) = -\frac{m_b -m_c}{q^2}\Big[m_{B_c}\Big(A_1(q^2) - A_2(q^2)\Big)+m_{J/\Psi}\Big(A_2(q^2) + A_1(q^2)-2\,A_0(q^2)\Big)\Big]\,.
\end{eqnarray}
A preliminary lattice calculation of $B_c \to J/\Psi$ transition form factors is reported in Ref.~\cite{Lytle:2016ixw,Colquhoun:2016osw}.

The $B \to D$ transition form factors $F_0(q^2)$ and $F_{+}(q^2)$, calculated using lattice QCD techniques, are collected from 
Ref.~\cite{Lattice:2015rga}. For the tensor form factor $F_T(q^2)$, we use Ref.~\cite{Melikhov:2000yu}. That is
\begin{eqnarray}
&&F_T(q^2)= \frac{0.69}{\Big(1-\frac{q^2}{6.4^2}\Big)\Big(1-0.56\frac{q^2}{6.4^2}\Big)}
\end{eqnarray}
Similarly, for $B \to D^{\ast}$ form factors, we follow the heavy quark effective theory~(HQET) approach of Ref.~\cite{Caprini:1997mu}. 
We refer to Ref.~\cite{Caprini:1997mu} for all the relevant equations.
 
We first perform a $\chi^2$ test to measure the disagreement of SM with the data. The $\chi^2$ is defined as
\begin{eqnarray}
\chi^2 = \sum_{i}\,\frac{\Big(\mathcal O_i^{\rm th} - \mathcal O_i^{\rm exp}\Big)^2}{\Big(\Delta\,\mathcal O_i^{\rm exp}\Big)^2}\,,
\end{eqnarray}
where $\mathcal O_i^{\rm exp}$ represents the measured central value of the observables and $\Delta\,\mathcal O_i^{\rm exp}$ represents 
corresponding $1\sigma$ uncertainty. Similarly, $\mathcal O_i^{\rm th}$ represents the theoretical prediction of the observables.
We include a total of three measurements for the evaluation of $\chi^2$, namely, $R_D$, $R_{D^{\ast}}$, and $R_{J/\Psi}$. We have not 
included $P_{\tau}^{D^{\ast}}$ in our fit as the error associated with it is rather large. We have found $\chi^2_{\rm min} = 16.3$ in 
the SM. The $\chi^2_{\rm min}$ in the SM is obtained by performing a random scan of all the theory input parameters such as CKM matrix
elements, meson decay constant, and meson to meson form factors within $1\sigma$ of their central values. The corresponding best estimates 
for all the observables are listed in Table.~\ref{tab2}. 

\begin{figure}[htbp]
\centering
\includegraphics[width=5.8cm,height=5.5cm]{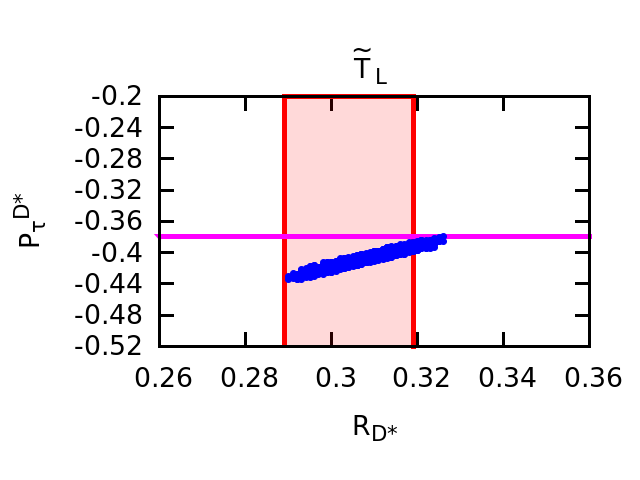}
\includegraphics[width=5.8cm,height=5.5cm]{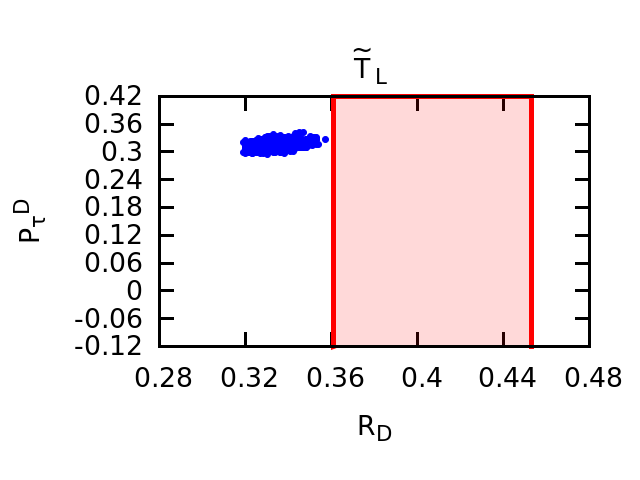}
\includegraphics[width=5.8cm,height=5.5cm]{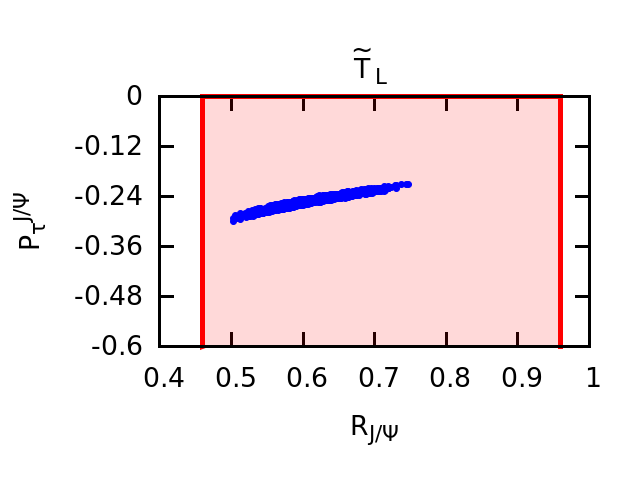}
\includegraphics[width=5.8cm,height=5.5cm]{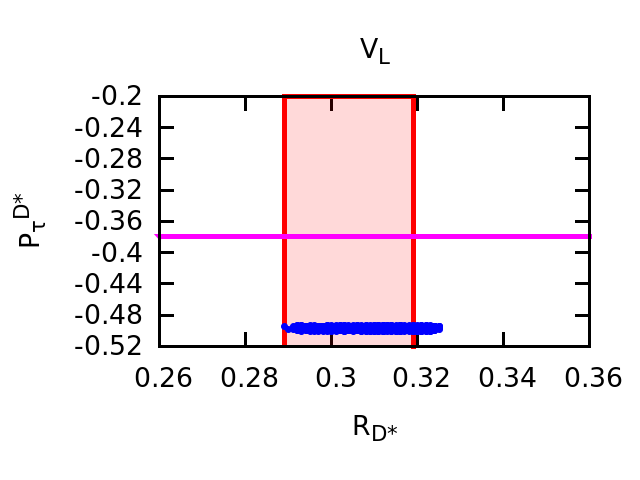}
\includegraphics[width=5.8cm,height=5.5cm]{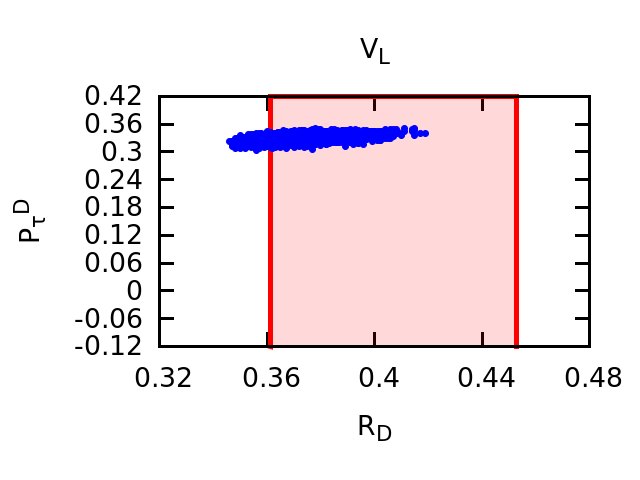}
\includegraphics[width=5.8cm,height=5.5cm]{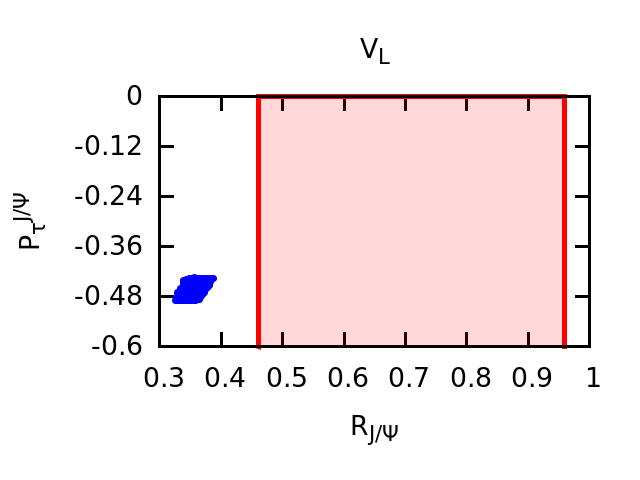}
\includegraphics[width=5.8cm,height=5.5cm]{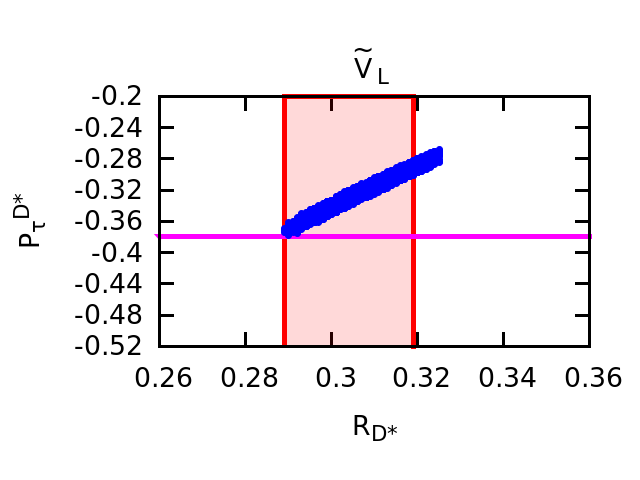}
\includegraphics[width=5.8cm,height=5.5cm]{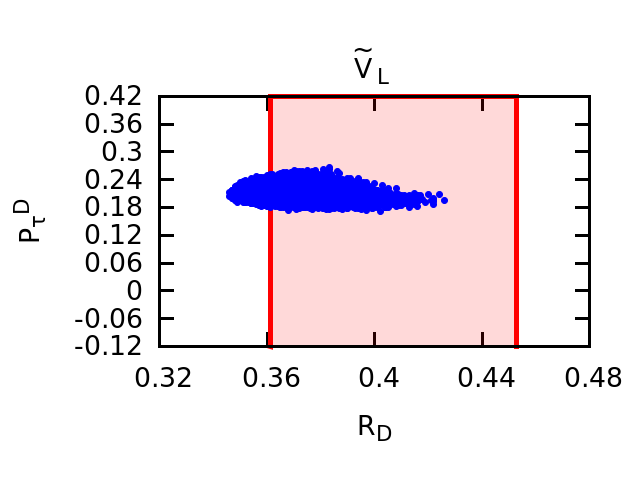}
\includegraphics[width=5.8cm,height=5.5cm]{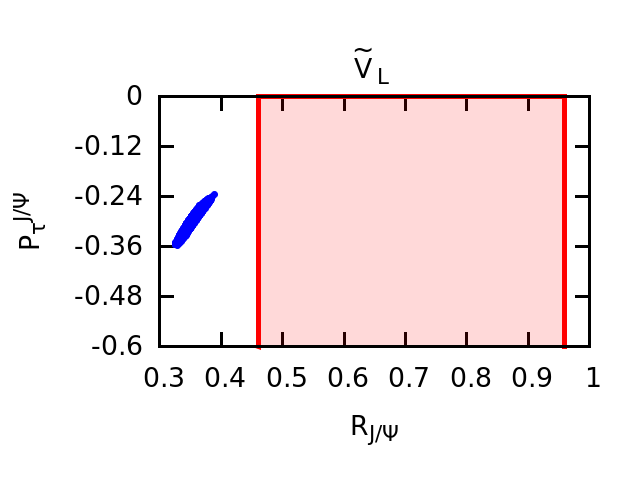}
\caption{The allowed
ranges in $(R_D,\,P_{\tau}^D)$, $(R_{D^{\ast}},\,P_{\tau}^{D^{\ast}})$, and
$(R_{J/\Psi},\,P_{\tau}^{J/\Psi})$ at $95\%$ CL are shown with blue bands.
The experimental $1\sigma$
range of $R_D$, $R_{D^{\ast}}$, and $R_{J/\Psi}$ are shown with light red bands. The horizintal line in the leftmost panel
represents the central value of $P_{\tau}^{D^{\ast}}$ reported by BELLE.}
\label{NP1}
\end{figure}

Let us now evaluate $\chi^2_{\rm min}$ for various NP scenarios. First we consider that NP contributions are coming from an operator 
characterized by a single NP Wilson Coefficient~(WC).
We have considered total $10$ such NP scenarios that involve left handed as well as right handed neutrino interactions. 
The $\chi^2_{\rm min}$ obtained in each scenario and the corresponding best estimates of all the observables are listed in Table.~\ref{tab2}. 
We observe that we obtain the best fit to the data with $\widetilde{T}_L$ NP coupling which corresponds to $\chi^2_{\rm min}=1.7$ followed
by $V_L$, $\widetilde{V}_L$, and $\widetilde{V}_R$ with $\chi^2_{\rm min}= 2.1$ each. The branching ratio of $B_c \to \tau\nu$ obtained in
each of these scenarios is consistent with $\mathcal B(B_c \to \tau\nu) \le 5\%$ obtained in the SM. With $\widetilde{T}_L$ NP coupling,
although the the best estimates of $R_{D^{\ast}}$ and $R_{J/\Psi}$ lie inside the $1\sigma$ experimental range, the best estimate
of $R_D$, however, lies outside $1\sigma$ of the experimental value. Similarly, with $V_L$ and $\widetilde{V}_{(L,\,R)}$ NP couplings
although, the best estimates of $R_D$ and $R_{D^{\ast}}$ lie inside the $1\sigma$ experimental range, the best estimate
of $R_{J/\Psi}$, however, lies outside $1\sigma$ of the experimental value.
\begin{table}[htbp]
\begin{ruledtabular}
\begin{tabular}{cccccccc}
Coefficients&Best fit value&$R_D$ &$ R_{D^{\ast}} $ & $R_{J/\Psi}$&$P_{\tau}^{D^{\ast}}$&$\mathcal B(B_c \to \tau\nu)\% $ &
$\chi^2_{\rm min}$ \\
\hline
SM& &$0.334$ &$0.255$ &$0.291$&$-0.501$&$2.3$& $16.3$ \\
$V_L$&$-2.11$&$0.398$ &$0.307$ &$0.356$ &$-0.494$ &$2.9$ &$2.1$ \\
$V_R$&$-0.09$&$0.276$ &$0.295$ &$0.345$ &$-0.496$ &$2.5$&$10.7$ \\
$S_L$&$-1.51$&$0.365$ &$0.330$ &$0.418$ &$-0.140$ &$121.5$&$5.2$ \\
$S_R$&$0.31$&$0.427$ &$0.266$ &$0.308$ &$-0.431$ &$12.1$&$9.6$ \\
$\widetilde{V}_L$&$0.48$&$0.398$ &$0.309$ &$0.367$ &$-0.311$ &$2.7$&$2.1$ \\
$\widetilde{V}_R$&$0.48$&$0.398$ &$0.309$ &$0.367$ &$-0.311$ &$2.7$&$2.1$ \\
$\widetilde{S}_L$&$0.73$&$0.432$ &$0.262$ &$0.302$ &$-0.513$ &$23.2$&$11.2$ \\
$\widetilde{S}_R$&$0.73$&$0.432$ &$0.262$ &$0.302$ &$-0.513$ &$23.2$&$11.2$ \\
$T_L$&$-0.08$&$0.309$ &$0.302$ &$0.457$ &$-0.467$ &$2.0$ &$5.6$ \\
$\widetilde{T}_L$&$0.27$&$0.352$ &$0.305$ &$0.585$ &$-0.412$ &$ 2.3$&$1.7$ \\
\end{tabular}
\caption{Best estimates of $R_D$, $R_{D^{\ast}}$, $R_{J/\Psi}$, $P_{\tau}^{D^{\ast}}$, and $\mathcal B(B_c \to \tau\nu)$ within the SM and 
within various NP scenarios.}
\label{tab2}
\end{ruledtabular}
\end{table}
We show in Fig.~\ref{NP1} the $95\%$ CL~(blue band) allowed ranges
in $(R_D,\,P_{\tau}^D)$, $(R_{D^{\ast}},\,P_{\tau}^{D^{\ast}})$, and $(R_{J/\Psi},\,P_{\tau}^{J/\Psi})$ for NP scenarios with $V_L$,
$\widetilde{V}_L$, and $\widetilde{T}_L$ NP couplings. We observe that, although, $V_L$ and $\widetilde{V}_L$ NP couplings 
can simulataneously explain the anomalies present in $R_D$ and $R_{D^{\ast}}$, these couplings, however, can not accommodate the
$R_{J/\Psi}$ data within $1\sigma$. We note that with $\widetilde{T}_L$ NP coupling, the range obtained for $R_{J/\Psi}$ lies within the
$1\sigma$ range of $R_{J/\Psi}$ data. However with $\widetilde{T}_L$ NP coupling, the range obtained for $R_D$ lies outside the $1\sigma$
experimental range. Again, we want to emphasise that the range in $\tau$ polarization fraction $P_{\tau}^{D^{\ast}}$ for each
scenarios lies inside the $1\sigma$ allowed range reported by Belle Colaboration. However, with $\widetilde{T}_L$ and $\widetilde{V}_L$
NP couplings, the central value lies inside the $95\%$~(blue) CL allowed bands.  
We notice that with $V_L$ NP coupling, the central value of $P_{\tau}^{D^{\ast}}$ reported by Belle 
Collaboration lies much above the range obtained in this scenario. 
However, the uncertainty associated with $P_{\tau}^{D^{\ast}}$ is rather large. Precise determination of
this parameter in the future will play a crucial role in identifying the exact nature of NP.

Now let us consider that NP contributions are coming from two different operators characterized by two NP WC.
We consider total $45$ such NP scenarios.
In Table.~\ref{tab3}, we show the best estimates of $R_D$, $R_{D^{\ast}}$, $R_{J/\Psi}$, $P_{\tau}^{D^{\ast}}$ and 
$\mathcal B(B_c \to \tau\nu)$ for each of these scenarios. We find that there are various NP scenarios that can simultaneously explain
$R_D$, $R_{D^{\ast}}$, and $R_{J/\Psi}$ anomalies. We notice that with
$(V_R,\,\widetilde{T}_L)$ NP couplings we obtain the best fit with the data with $\chi^2_{\rm min} = 0.43\times 10^{-2}$ 
followed by $(\widetilde{V}_R,\,\widetilde{T}_L$), $(S_R,\,\widetilde{T}_L)$, and $(\widetilde{V}_L,\,\widetilde{T}_L)$ NP couplings 
with $\chi^2_{\rm min} =0.013,\,0.026,\,{\rm and}\, 0.11$,
respectively. However, with $(S_R,\,\widetilde{T}_L)$ NP couplings, the best estimates of $\mathcal B(B_c \to \tau\nu)=80.6\%$ is much
above the upper bound of $\mathcal B(B_c \to \tau\nu) \le 30\%$ estimated in the SM.  Hence although, we get a much better fit with 
this NP structure, $(S_R,\,\widetilde{T}_L)$ NP couplings can not accommodate the $B_c \to \tau\nu$ data.
\begin{table}[htbp]
\begin{ruledtabular}
\begin{tabular}{cccccccc}
Coefficients&Best fit value&$R_D$ &$ R_{D^{\ast}} $ & $R_{J/\Psi}$&$P_{\tau}^{D^{\ast}}$&$\mathcal B(B_c \to \tau\nu)\% $ &$\chi^2_{\rm min}$ \\
\hline
$(V_L,\,V_R)$&$(-2.14,\,-0.04)$&$0.406$ &$0.308$ &$0.357$ &$-0.500$ &$2.7$ &$2.1$ \\
$(V_L,\,S_L)$&$(-1.94,\,1.62)$&$0.403$ &$0.308$ &$0.383$ &$-0.092$ &$137.3$ &$1.8$ \\
$(V_L,\,S_R)$&$(0.09,\,0.13)$&$0.411$ &$0.305$ &$0.352$ &$-0.475$ &$6.3$ &$2.1$ \\
$(V_L,\,T_L)$&$(-2.13,\,-0.03)$&$0.404$ &$0.302$ &$0.320$ &$-0.502$ &$2.8$ &$2.5$ \\
$(V_L,\,\widetilde{V}_L)$&$(-1.68,\,-0.87)$&$0.397$ &$0.306$ &$0.356$ &$0.121$ &$2.7$ &$2.1$ \\
$(V_L,\,\widetilde{V}_R)$&$(-1.68,\,-0.87)$&$0.397$ &$0.306$ &$0.356$ &$0.121$ &$2.7$ &$2.1$ \\
$(V_L,\,\widetilde{S}_L)$&$(-2.10,\,-0.29)$&$0.403$ &$0.305$ &$0.353$ &$-0.498$ &$5.9$ &$2.1$ \\
$(V_L,\,\widetilde{S}_R)$&$(-2.10,\,-0.29)$&$0.403$ &$0.305$ &$0.353$ &$-0.498$ &$5.9$ &$2.1$ \\
$(V_L,\,\widetilde{T}_L)$&$(0.04,\,-0.23)$&$0.361$ &$0.312$ &$0.534$ &$-0.439$ &$2.3$ &$1.8$ \\[0.1cm]
$(V_R,\,S_L)$&$( 0.08,\,-1.68)$&$0.406$ &$0.307$ &$0.372$ &$-0.070$ &$154.1$ &$1.9$ \\
$(V_R,\,S_R)$&$(-0.09,\,0.27)$&$0.402$ &$0.306$ &$0.359$ &$-0.445$ &$10.5$ &$2.1$ \\
$(V_R,\,T_L)$&$(0.25,\,-0.27)$&$0.399$ &$0.304$ &$0.822$ &$-0.365$ &$1.2$ &$0.24$ \\
$(V_R,\,\widetilde{V}_L)$&$(0.06,\,0.56)$&$0.408$ &$0.303$ &$0.355$ &$-0.238$ &$2.6$ &$2.1$ \\
$(V_R,\,\widetilde{V}_R)$&$(0.06,\,0.56)$&$0.408$ &$0.303$ &$0.355$ &$-0.238$ &$2.6$ &$2.1$ \\
$(V_R,\,\widetilde{S}_L)$&$(-0.10,\,0.72)$&$0.410$ &$0.306$ &$0.359$ &$-0.507$ &$22.9$ &$2.0$ \\
$(V_R,\,\widetilde{S}_R)$&$(-0.10,\,0.72)$&$0.410$ &$0.306$ &$0.359$ &$-0.507$ &$22.9$ &$2.0$ \\
$\bf (V_R,\,\widetilde{T}_L)$&$(0.08,\,0.33)$&$0.405$ &$0.304$ &$0.722$ &$-0.365$ &$1.9$ &$0.43\times 10^{-2}$ \\[0.1cm]
$(S_L,\,S_R)$&$(-0.49,\,0.71)$&$0.405$ &$0.309$ &$0.386$ &$-0.224$ &$79.9$&$1.8$ \\
$(S_L,\,T_L)$&$(0.20,\,-0.10)$&$0.413$ &$0.303$ &$0.512$ &$-0.486$ &$0.04$&$0.66$ \\
$(S_L,\,\widetilde{V}_L)$&$(0.06,\,-0.47)$&$0.400$ &$0.305$ &$0.358$ &$-0.326$ &$1.5$&$2.0$ \\
$(S_L,\,\widetilde{V}_R)$&$(0.06,\,-0.47)$&$0.400$ &$0.305$ &$0.358$ &$-0.326$ &$1.5$&$2.0$ \\
$(S_L,\,\widetilde{S}_L)$&$(-1.08,\,-0.86)$&$0.407$ &$0.311$ &$0.393$ &$-0.271$ &$109.1$&$1.8$ \\
$(S_L,\,\widetilde{S}_R)$&$(-1.08,\,-0.86)$&$0.407$ &$0.311$ &$0.393$ &$-0.271$ &$109.1$&$1.8$ \\
$(S_L,\,\widetilde{T}_L)$&$(0.18,\,-0.28)$&$0.410$ &$0.305$ &$0.629$ &$-0.429$ &$0.10$&$0.12$ \\[0.1cm]
$(S_R,\,T_L)$&$(-1.66,\,-0.12)$&$0.421$ &$0.307$ &$0.548$ &$-0.544$ &$91.2$&$0.57$ \\
$(S_R,\,\widetilde{V}_L) $&$(0.14,\,-0.44)$&$0.404$ &$0.305$ &$0.360$ &$-0.317$ &$6.4$&$2.0$ \\
$(S_R,\,\widetilde{V}_R) $&$(0.14,\,-0.44)$&$0.404$ &$0.305$ &$0.360$ &$-0.317$ &$6.4$&$2.0$ \\
$(S_R,\,\widetilde{S}_L)$&$(0.30,\,0.10)$&$0.438$ &$0.266$ &$0.307$ &$-0.438$ &$12.7$&$9.8$ \\
$(S_R,\,\widetilde{S}_R)$&$(0.30,\,0.10)$&$0.438$ &$0.266$ &$0.307$ &$-0.438$ &$12.7$&$9.8$ \\
$(S_R,\,\widetilde{T}_L)$&$(-1.63,\,-0.31)$&$0.405$ &$0.305$ &$0.676$ &$-0.481$ &$80.5$&$0.26\times 10^{-1}$ \\[0.1cm]
$(T_L,\,\widetilde{V}_L)$&$(0.01,\,0.48)$&$0.394$ &$0.308$ &$0.345$ &$-0.310$ &$2.5$&$2.3$ \\
$(T_L,\,\widetilde{V}_R)$&$(0.01,\,0.48)$&$0.394$ &$0.308$ &$0.345$ &$-0.310$ &$2.5$&$2.3$ \\
$(T_L,\,\widetilde{S}_L)$&$(-0.08,\,-0.66)$&$0.405$ &$0.302$ &$0.472$ &$-0.473$ &$18.1$&$0.94$ \\
$(T_L,\,\widetilde{S}_R)$&$(-0.08,\,-0.66)$&$0.405$ &$0.302$ &$0.472$ &$-0.473$ &$18.1$&$0.94$ \\
$(T_L,\,\widetilde{T}_L)$&$(0.12,\,-0.38)$&$0.392$ &$ 0.307$ &$0.741$ &$-0.345$ &$2.3$&$0.16$ \\[0.1cm]
$(\widetilde{V}_L,\,\widetilde{V}_R)$&$(-0.05,\,-0.52)$&$0.413$ &$0.308$ &$0.362$ &$-0.313$ &$2.5$&$2.1$ \\
$(\widetilde{V}_L,\,\widetilde{S}_L)$&$(0.37,\,-0.80)$&$0.412$ &$0.306$ &$0.364$ &$-0.415$ &$34.2$&$2.0$ \\
$(\widetilde{V}_L,\,\widetilde{S}_R)$&$(0.44,\,0.15)$&$0.405$ &$0.304$ &$0.358$ &$-0.342$ &$ 4.9$&$2.0$ \\
$\bf (\widetilde{V}_L,\,\widetilde{T}_L)$&$(-0.34,\,-0.35)$&$0.405$ &$0.306$ &$0.635$ &$-0.427$ &$2.6$&$0.11$ \\[0.1cm]
$(\widetilde{V}_R,\,\widetilde{S}_L)$&$(0.44,\,0.15)$&$0.405$ &$0.304$ &$0.358$ &$-0.342$ &$4.9$&$2.0$ \\
$(\widetilde{V}_R,\,\widetilde{S}_R)$&$(0.37,\,-0.80)$&$0.412$ &$0.306$ &$0.364$ &$-0.415$ &$34.2$&$2.0$ \\
$\bf (\widetilde{V}_R,\,\widetilde{T}_L)$&$(-0.68,\,0.44)$&$0.405$ &$0.304$ &$0.686$ &$-0.446$ &$3.1$&$0.13\times 10^{-1}$ \\[0.1cm]
$(\widetilde{S}_L,\,\widetilde{S}_R)$&$(1.36,\,-0.73)$&$0.405$ &$0.305$ &$0.374$ &$-0.586$ &$206.5$&$1.9$ \\
$(\widetilde{S}_L,\,\widetilde{T}_L)$&$(0.49,\,-0.27)$&$0.399$ &$0.307$ &$0.619$ &$-0.418$ &$12.1$&$0.22$ \\[0.1cm]
$(\widetilde{S}_R,\,\widetilde{T}_L)$&$(0.49,\,-0.27)$&$0.399$ &$0.307$ &$0.619$ &$-0.418$ &$12.1$&$0.22$ \\[0.1cm]
\end{tabular}
\caption{Best estimates of $R_D$, $R_{D^{\ast}}$, $R_{J/\Psi}$, $P_{\tau}^{D^{\ast}}$, and $\mathcal B(B_c \to \tau\nu)$ 
within various NP scenarios.}
\label{tab3}
\end{ruledtabular}
\end{table}

\begin{figure}[htbp]
\centering
\includegraphics[width=4.3cm,height=4.3cm]{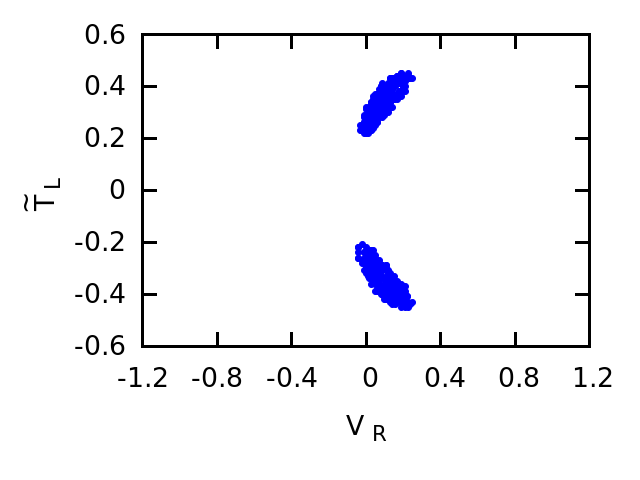}
\includegraphics[width=4.3cm,height=4.3cm]{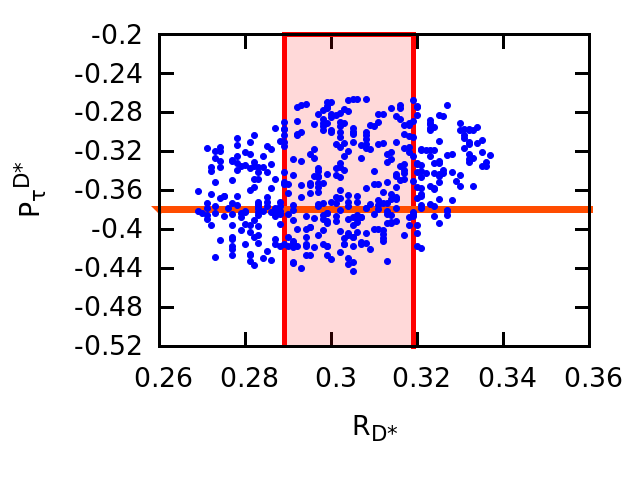}
\includegraphics[width=4.3cm,height=4.3cm]{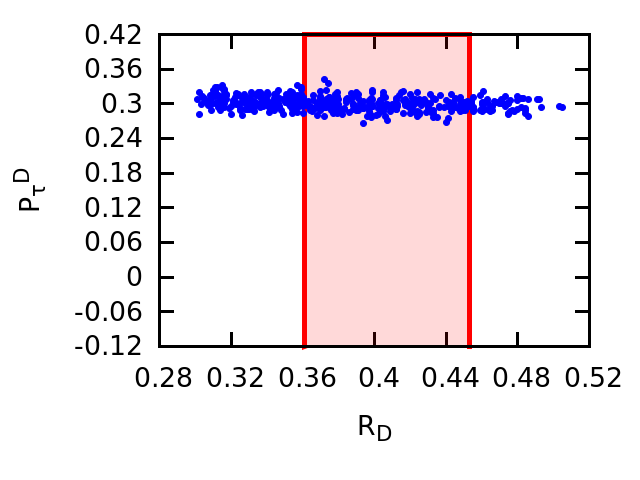}
\includegraphics[width=4.3cm,height=4.3cm]{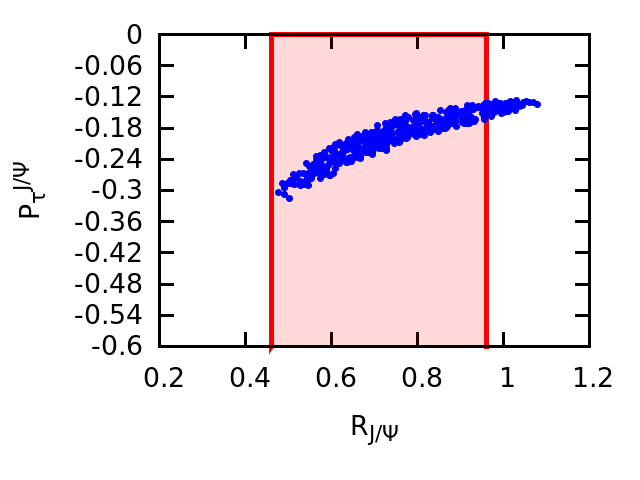}
\includegraphics[width=4.3cm,height=4.3cm]{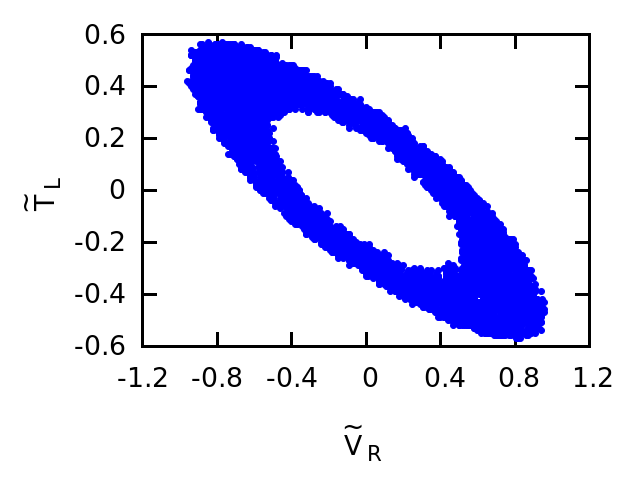}
\includegraphics[width=4.3cm,height=4.3cm]{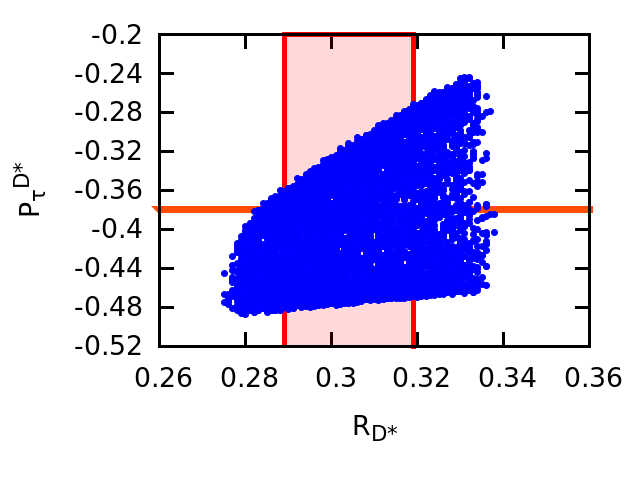}
\includegraphics[width=4.3cm,height=4.3cm]{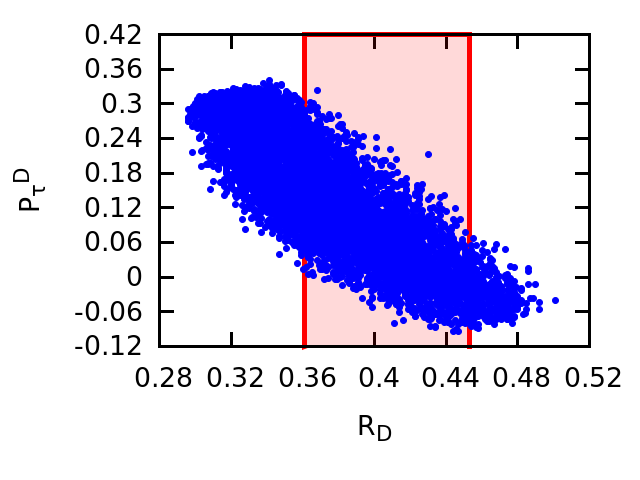}
\includegraphics[width=4.3cm,height=4.3cm]{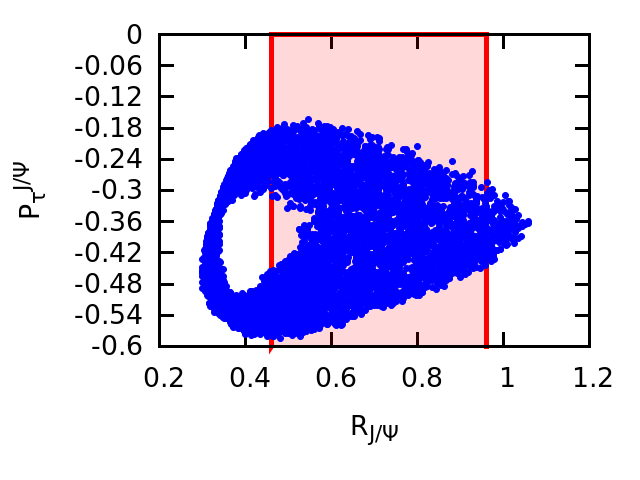}
\includegraphics[width=4.3cm,height=4.3cm]{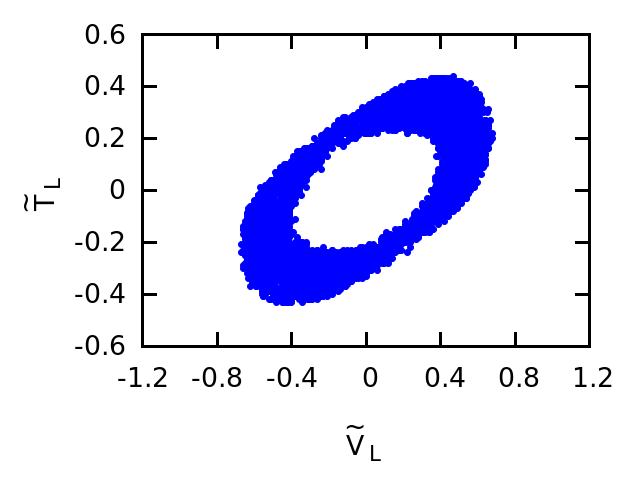}
\includegraphics[width=4.3cm,height=4.3cm]{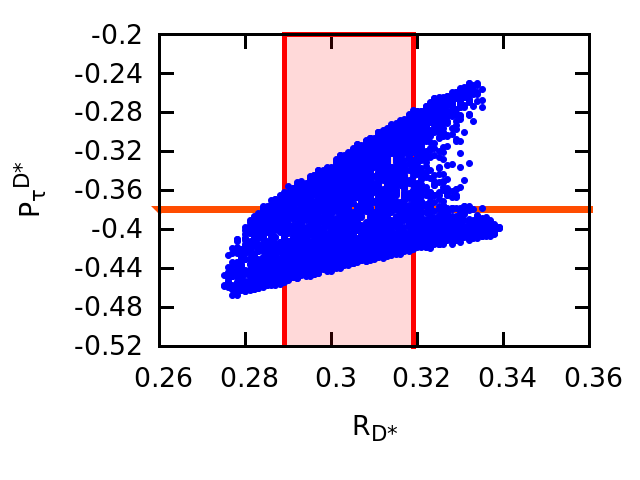}
\includegraphics[width=4.3cm,height=4.3cm]{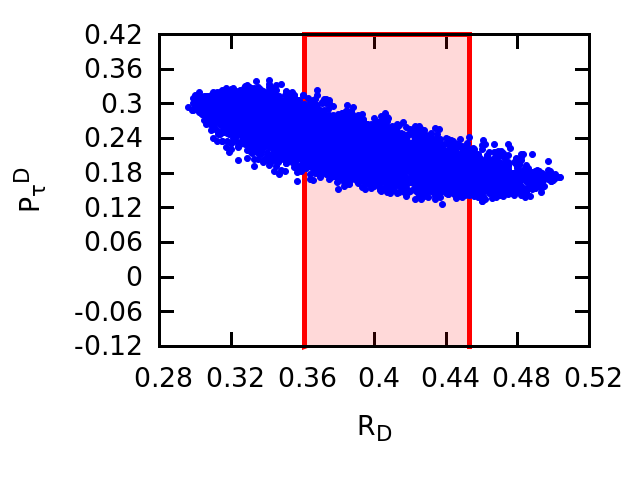}
\includegraphics[width=4.3cm,height=4.3cm]{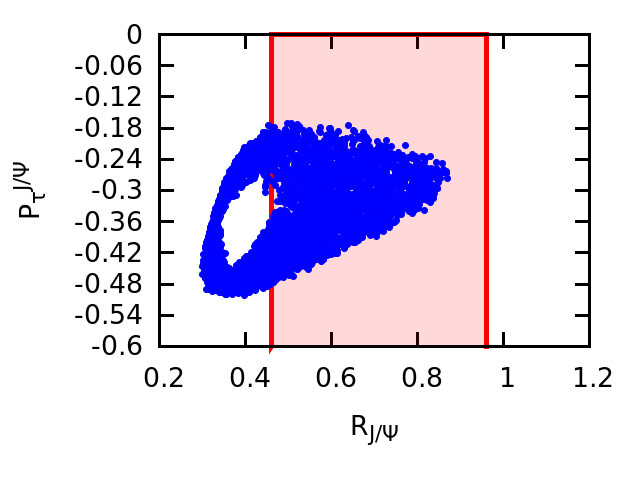}
\caption{The allowed NP parameter space~(leftmost panels) at $95\%$~(blue) CL from $R_D$, $R_{D^{\ast}}$, and $R_{J/\Psi}$
measurements. The corresponding allowed
ranges in $(R_D,\,P_{\tau}^D)$, $(R_{D^{\ast}},\,P_{\tau}^{D^{\ast}})$, and
$(R_{J/\Psi},\,P_{\tau}^{J/\Psi})$ are shown with blue dots.
The experimental $1\sigma$
range of $R_D$, $R_{D^{\ast}}$, and $R_{J/\Psi}$ are shown with light red bands. The horizintal line in the second panel
represents the central value of $P_{\tau}^{D^{\ast}}$ reported by BELLE.}
\label{NP2}
\end{figure}

We show in Fig.~\ref{NP2} the allowed range in $(V_R,\,\widetilde{T}_L)$, $(\widetilde{V}_R,\,\widetilde{T}_L)$, and 
$(\widetilde{V}_L,\,\widetilde{T}_L)$ NP parameter space at $95\%$~(blue) CL obtained from the measured values of $R_D$, 
$R_{D^{\ast}}$, and $R_{J/\Psi}$. We also show the allowed ranges in $(R_D,\,P_{\tau}^D)$, $(R_{D^{\ast}},\,P_{\tau}^{D^{\ast}})$, and 
$(R_{J/\Psi},\,P_{\tau}^{J/\Psi})$ for each of these NP scenarios.
It is clear that with all these NP couplings, the
allowed ranges in $R_D$, $R_{D^{\ast}}$,$R_{J/\Psi}$, and $P_{\tau}^{D^{\ast}}$ at $95\%$~(blue) CL do overlap with the 
experimentally allowed values within $1\sigma$. It should, however, be mentioned that with $(\widetilde{V}_R,\,\widetilde{T}_L)$ NP
couplings, the value of $P_{\tau}^{D}$ can be negative depending on the value of the NP couplings. Similarly, we do not see much variation
of $P_{\tau}^{D}$ with $(V_R,\,\widetilde{T}_L)$ NP couplings. Measurement of $P_{\tau}^{D}$ and $P_{\tau}^{J/\Psi}$ in future will also
play a crucial role.

\section{conclusion}
\label{con}
In view of the recent result of $R_{J/\Psi}$, we investigate $R_D$, $R_{D^{\ast}}$, and $R_{J/\Psi}$ anomalies using a model-independent 
framework. We consider total $55$ NP models consisting of single as well as double NP coefficients and evaluate $\chi^2_{\rm min}$ by 
including three measurements, namely, $R_D$, $R_{D^{\ast}}$, and $R_{J/\Psi}$. We find various NP models that can simultaneously explain
$R_D$, $R_{D^{\ast}}$, and $R_{J/\Psi}$ anomalies within $1\sigma$.
We also give prediction on the $\tau$ polarization fraction parameter for all these decay modes in each NP scenarios. It should 
be noted that
a more precise measurement on $P_{\tau}^{D^{\ast}}$ in future will be crucial to identify the 
true nature of NP. Again, in view of the immense importance of $R_D$, $R_{D^{\ast}}$, and $R_{J/\Psi}$, both experimentally and 
theoretically, it is important to ensure that theoretical calculations of various form factors are very precise. At the same time, 
measurement of $P_{\tau}^D$ and $P_{\tau}^{J/\Psi}$ in future will be crucial to rule out various NP scenarios.

\end{document}